\documentclass{article}
\usepackage[utf8]{inputenc}
\usepackage{import}
\usepackage{amsmath}
\usepackage{slashed}
\usepackage{mathtools}
\usepackage{caption}
\usepackage{authblk}
\usepackage{blindtext}
\usepackage{subcaption}
\usepackage{color}
\usepackage{setspace}

\DeclarePairedDelimiterX\braket[2]{\langle}{\rangle}{#1 \delimsize\vert #2}
\usepackage{abstract}
\usepackage{cite}

\usepackage{graphicx}
\usepackage{multirow}
\title{\textbf{More on Denominator Regularization in Quantum Field Theory}}
\author[1]{Anshika Bansal\thanks{ anshika@prl.res.in}}
\author[1]{Namit Mahajan\thanks{ nmahajan@prl.res.in}}
\author[1,2]{Dayanand Mishra\thanks{dayanand@prl.res.in}}
\affil[1]{Physical Research Laboratory, Ahmedabad, 380009, India.}
\affil[2]{Indian Institute of Technology, Gandhinagar, 382424, India.}
\date{}
\begin{document}
\maketitle
\doublespacing
\begin{abstract}
Recently \cite{Horowitz:2022rpp,Horowitz:2022uak}, denominator regularisation (Den. Reg.) scheme has been proposed to handle divergences in quantum field theory. It is shown to yield results as simple as in dimensional regularisation scheme and also shown to be compatible with minimal subtraction scheme by taking several examples in field theory. In this article, we point out some interesting features of this regularisation scheme which appear as a consequence of the requirement of gauge invariance by revisiting QED vacuum polarisation and $H\to\gamma\gamma$ decay in this regularisation scheme.
\end{abstract}
\section{Introduction}
In quantum field theory, the higher order contributions (the loop diagrams) often lead to infinities. To cure these infinities, one picks up a regularisation scheme, followed by renormalisation to finally get physical, finite answers. The most commonly used regularisation scheme is dimensional regularisation \cite{tHooft:1973mfk,Ashmore:1972uj,Bollini:1972bi} as it preserves the symmetries of the theory. Other regularisation schemes often used are the cut-off regularisation \cite{Heisenberg:1938xla}, Pauli Villars regularisation \cite{Pauli:1949zm}, etc. (See \cite{Schwinger:1951nm,Freedman:1991tk,Haagensen:1992am,Reisz:1987da} for other popular regularisation schemes). Recently, in \cite{Horowitz:2022uak,Horowitz:2022rpp}, a new novel regularisation scheme known as denominator regularisation has been proposed. The authors claimed that it is as simple to apply as dimensional regularisation and is compatible with a minimal subtraction scheme: it preserves Lorentz invariance, gauge invariance, and supersymmetry, and correctly predicts the axial anomaly as well as yields the Green's function that satisfies the Callan-Symanzik equation. Denominator regularisation is claimed to have several other applications such as in asymmetric spacetimes, curved spacetimes, and thermal field theory.

In this note, we consider the amplitude of the Higgs boson decaying to two photons. Den. Reg. being a gauge invariance preserving scheme is expected to produce correct results. The aim is to check the applicability of this scheme to such a case and see if any modifications are needed beyond what is the suggested prescription. We will first discuss the structure of the amplitude in brief and then compute this decay amplitude in the newly prescribed regualrisation scheme.

 The rest of the article is arranged as follows. In Section-\ref{htogg}, we start by discussing the Higgs decay to two photons in dimensional regularisation and identify the main issue of concern. In Section-\ref{htoggdr}, we will discuss how the amplitude looks in the newly proposed Den. Reg. scheme, and discuss some interesting features which appear as a consequence of the requirement of gauge invariance in QED. Finally, we close with a brief discussion.

\section{Higgs decay to two photons}
\label{htogg}
The Higgs boson was discovered via its decay into two photons \cite{CMS:2012qbp,ATLAS:2012yve}. This decay is possible only via the quantum loop corrections induced by the W-boson and the fermions, mainly the top quark, because the photon is massless and can not directly couple to the Higgs boson. 
\begin{figure}
    \centering
    \begin{subfigure}{0.32\textwidth}
    \includegraphics[width=0.9\textwidth]{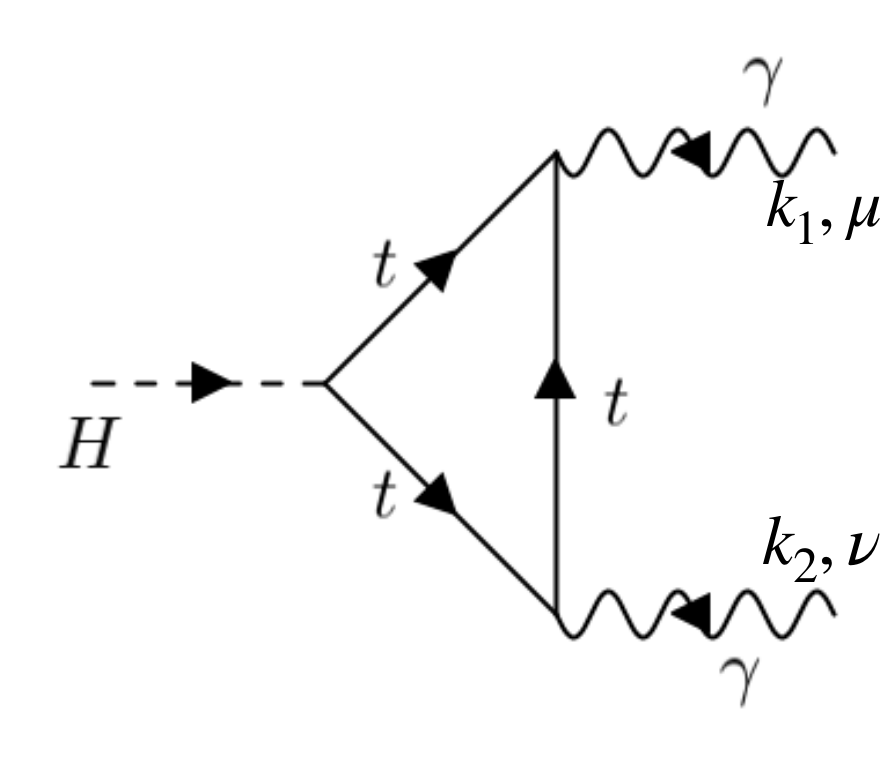}
    \end{subfigure}
    \begin{subfigure}{0.32\textwidth}
    \includegraphics[width=0.9\textwidth]{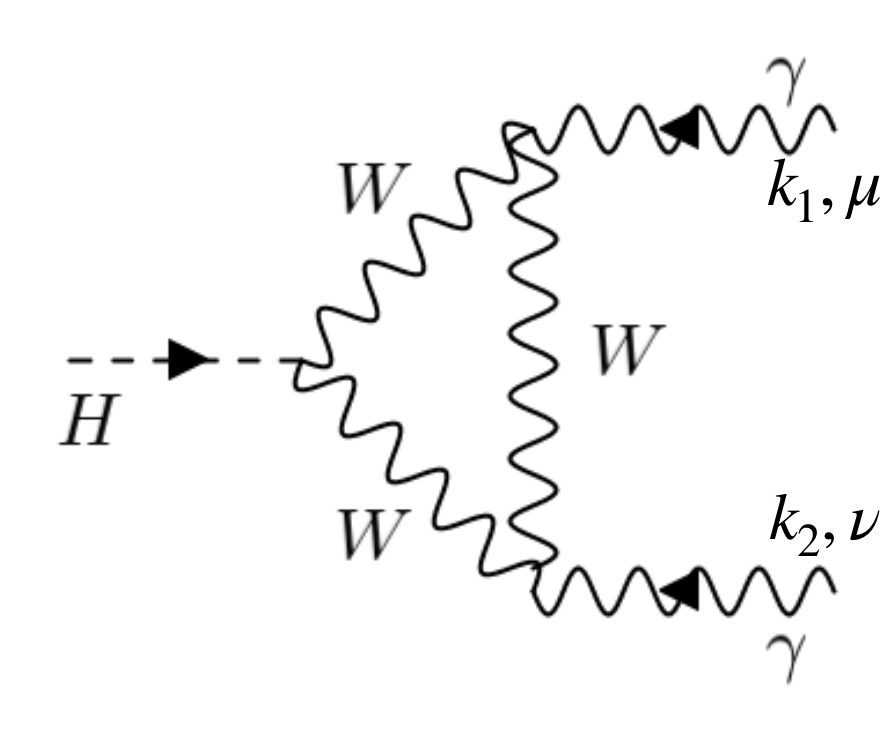}
    \end{subfigure}
    \begin{subfigure}{0.32\textwidth}
    \includegraphics[width=0.9\textwidth]{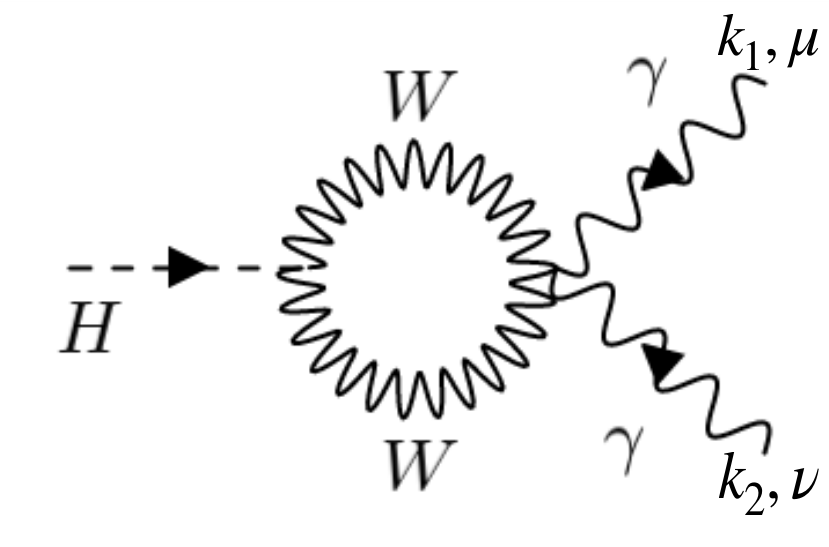}
    \end{subfigure}
    \caption{Feynman diagrams showing the fermion (top quark) and W-boson loop contribution to $H\to \gamma \gamma$.}
    \label{htoggp1}
\end{figure}
The contribution to the amplitude of
$H\to \gamma \gamma$ coming from the fermion loop reads as
\begin{eqnarray}
	\mathcal{M}&=&\int \frac{d^4 p}{(2\pi)^4}\frac{i^5 e^2 Tr\left[\left(\slashed{p}-\slashed{k_1}+m_q\right) \gamma^{\mu}\left(\slashed{p}-\slashed{k_1}+m_q\right)\gamma^{\nu}\left(\slashed{p}+\slashed{k_2}+m_q\right)\right]}{\left((p-k_1)^2-m_{q}^2\right)\left(p^2-m_q^2\right)\left((p+k_2)^2-m_q^2\right)}\nonumber \\
	&=& \int_{0}^{1}dx \int_{0}^{1-x} dz \int \frac{d^4 p}{(2\pi)^4} \frac{ie^2 8m_q\left[(4p^{\mu}p^{\nu}-p^2 g^{\mu\nu})+2(p^{\mu}k_2^{\nu}-p^{\nu}k_2^{\mu})-k_1^{\mu}k_2^{\nu}+k_1^{\nu}k_2^{\mu}+(m_q^2 -k_1. k_2) g^{\mu \nu}\right]}{\left((p+z k_2-x k_1)^2+x z m_H^2 -m_q^2\right)^3}\nonumber\\
    &=& \int_{0}^{1}dx \int_{0}^{1-x} dz \int \frac{d^4 l}{(2\pi)^4} i e^2 4 m_q \Big[ \frac{2\left(4 l^{\mu}l^{\nu}-l^2 g^{\mu \nu}+(4z^2-2z) k_3^{\mu}k_3^{\nu}+(4x^2-2x)k_2^{\mu}k_2^{\nu} \right)}{(l^2-\Delta)^3}\nonumber\\&+&\frac{2(2x+2z-4zx)k_2^{\mu}k_3^{\nu}+2 x z m_H^2 g^{\mu \nu}-2k_2^{\nu}k_3^{\mu}+2(m_q^2-\frac{1}{2}m_H^2) g^{\mu \nu} }{(l^2-\Delta)^3}\Big]
	\end{eqnarray}
 where, $k_1$ and $k_2$ are the momenta of the final state photons and the quantity $\Delta$ is a scalar quantity which is a function of $k_1$, $k_2$ and the Feynman parameters $x$ and $z$.
 
The interesting part of this amplitude is the integral term represented as	
	\begin{eqnarray}
	    I^{\mu\nu}=\int \frac{d^4 l}{(2\pi)^4} \Big[\frac{4l^{\mu}l^{\nu}}{(l^2-\Delta)^3}-\frac{l^2 g^{\mu \nu}}{(l^2-\Delta)^3} \Big]
	    \label{eqn_int}
	\end{eqnarray}
The contribution stemming from W-boson loops is similar but more illuminating. In a series of papers \cite{Gastmans:2011ks, Gastmans:2011wh}, it was claimed that the earlier calculations of $H\to \gamma \gamma$ amplitude were erroneous and the source of the trouble was the employment of dimensional regularisation. The authors in \cite{Gastmans:2011ks,Gastmans:2011wh} had mistakenly canceled the two terms in Eq. (2) by a naive use of symmetric integration in four space-time dimensions. The essential mistake was ignoring the fact that both these integrals are divergent by themselves and require regularisation before proceeding further. In dimensional regularisation, as pointed out by \cite{Marciano:2011gm, Shifman:2011ri,Huang:2011yf,Jegerlehner:2011jm}, the symmetric integration yields
	\begin{eqnarray}
	    l^{\mu}l^{\nu}=\frac{l^2}{d} g^{\mu\nu}.
	\end{eqnarray}
Therefore, in dimensional regularisation, it is very clear that the above integral will not be zero and reads as 
	\begin{eqnarray}
	    I^{\mu\nu}&=&g^{\mu \nu}(\frac{4}{d}-1)\int \frac{d^d l}{(2\pi)^d} \frac{l^2}{(l^2-\Delta)^3}\nonumber\\
	    &=&g^{\mu \nu}(\frac{4}{d}-1) \frac{i}{(4\pi)^{d/2}}\frac{d}{2}\frac{\Gamma(2-\frac{d}{2})}{\Gamma(3)}\left(\frac{1}{\Delta}\right)^{2-\frac{d}{2}}
	    \label{eqn_pes}
	\end{eqnarray}
which for $d=4-\epsilon$ results into 
\begin{eqnarray}
	     I^{\mu\nu}&=&g^{\mu \nu}\frac{\epsilon}{2}\frac{i}{(4\pi)^2}\left(\frac{2}{\epsilon}+\mathcal{O}(\epsilon)\right)\nonumber\\
	     &=& \frac{i}{32\pi^2}g^{\mu\nu}.
	     \label{int_res}
\end{eqnarray}
For the W-boson contribution, one finally has the amplitude \cite{Marciano:2011gm}
\begin{align}
    \mathcal{M}_{W}= \frac{e^2 g}{(4\pi)^2 m_W} \left[2+3\beta +3\beta(2-\beta) f(\beta)\right] \left(k_1.k_2 g^{\mu\nu} -k_2^\mu k_1^\nu\right) \epsilon_\mu(k_1) \epsilon_\nu(k_2)
\end{align}
where, $\beta= \frac{4m_W^2}{m_H^2}$ and 
\begin{equation}
    f(\beta) = \begin{cases}
    \text{arcsin}^2\left(\beta^{-\frac{1}{2}}\right)  &\text{for } \beta \geq 1 \\
    -\frac{1}{4} \left[\text{ln} \frac{1+\sqrt{1-\beta}}{1-\sqrt{1-\beta}}-i\pi\right]^2 & \text{for } \beta <1
    \end{cases}.
\end{equation}
In the limit $m_{H}\to \infty$, the fermion contribution drops out. The same does not happen for the W-contribution. This non-decoupling is related to the fact that the longitudinal polarization of gauge bosons, $\epsilon^\mu_L\sim \frac{p^\mu}{m_W}$, yielding a non-zero $H\to \gamma\gamma$ via longitudinal W's. This is captured by the first term in the amplitude above (term independent of $\frac{m_W}{m_H}$). $I^{\mu\nu}$ as obtained in Eqn.(\ref{int_res}) is crucial for this purpose.

We thus see that $H\to \gamma\gamma$ can become an important litmus test of a regularisation scheme. We next apply it to the denominator regularisation scheme.

\section{Denominator regularisation and the Higgs decay}
\label{htoggdr}
In \cite{Horowitz:2022uak,Horowitz:2022rpp}, the authors have proposed Den. Reg. scheme as a possible alternative to the usually adopted dimensional regularization scheme. In the newly proposed scheme, one first combines the denominators using Feynman parameters, very much like in dimensional regularisation, and then analytically continues the power, '$n$', of the single combined denominator from $n$ to $n+\epsilon$, where $\epsilon$ is chosen to be sufficiently large in order to ensure that the integral is UV convergent. Like dimensional regularisation, this method also introduces a fictitious scale $\mu$ to preserve the dimensionality of the Green's function and eventually satisfy the Callan-Symanzik equation. Other than this fictitious scale, there is another crucial ingredient for the completion of this regularisation: an overall coefficient function $f_{(n,p)}$ which smoothly goes to unity as $\epsilon\to 0$. $n$ and $p$ here denote the original power of the denominator (after employing Feynman parametrization) and the superficial degree of divergence of the integral, respectively. This function can be uniquely fixed \cite{Horowitz:2022uak} by ``\textit{minimally requiring that the Laurent expansion of the amplitude in $\epsilon$ has only a simple pole at $\epsilon=0$, including in the massless limit"}.

This regularisation scheme is found to correctly predict the axial anomaly. It preserves the symmetries of the theory as well like gauge invariance, Lorentz invariance, supersymmetry, etc. Now, as this regularisation is supposed to preserve gauge invariance, reproducing the $H\to \gamma \gamma$ amplitude in this scheme and comparing it with the results obtained from dimensional regularisation scheme can shed more light on the scheme itself, and thus becomes a crucial test for this scheme. In order to do that, we will now focus only on the integral in Eqn.(\ref{eqn_int}) as that is essentially the non-trivial part of the amplitude as discussed above. 

In Den. Reg., the integral in Eqn.(\ref{eqn_int}) will be modified as 
	\begin{eqnarray}
	    I^{\mu\nu}= \int \frac{d^4l}{(2\pi)^4} \left[\frac{4l^\mu l^\nu}{(l^2-\Delta)^3} - \frac{l^2 g^{\mu\nu}}{(l^2-\Delta)^3}\right]\underset{\text{Den. reg.}}{\xrightarrow{\hspace*{1.2cm}}}  \mu^{2\epsilon} f_{(3,0)} \int \frac{d^4l}{(2\pi)^4}\left[\frac{4l^\mu l^\nu}{(l^2-\Delta)^{3+\epsilon}} - \frac{l^2 g^{\mu\nu}}{(l^2-\Delta)^{3+\epsilon}}\right]
	    \label{eqn_imunuf}
	\end{eqnarray}

\begin{figure}
    \centering    \includegraphics[width=0.25
\textwidth]{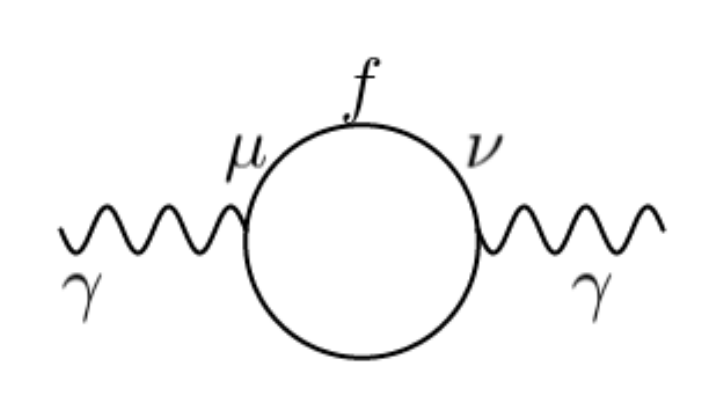}
    \caption{Feynman diagram for QED vacuum polarization.}
    \label{vac-pol}
\end{figure}

Following the prescription given in \cite{Horowitz:2022uak}, both these terms will have a common function $f_{(3,0)}$, which then is just an overall factor outside the integral. Further, $ l^{\mu}l^{\nu}=\frac{l^2}{4} g^{\mu\nu}$, since one now directly deals with a four dimensional integral. As a result, the numerator is trivially zero after regulating the integral via denominator regularisation. Therefore, Den. Reg. in the present form seems to fail in obtaining the desired result for $H\to \gamma\gamma$, since now $I^{\mu\nu}$ as in Eqn.(\ref{int_res}) is simply absent. The question then arises whether it is really failing in explaining the $H\to \gamma\gamma$ decay or we are missing something important which is leading to erroneous results. In the quest for the answer to this puzzling issue, let us first revisit the QED vacuum polarisation which had been already discussed in \cite{Horowitz:2022uak}. The amplitude for QED vacuum polarisation is given by (see Fig. (\ref{vac-pol}))
\begin{eqnarray}
     \pi_{\alpha \beta}&=&(\frac{-i g_{\alpha \mu}}{p^2})i\Pi^{\mu \nu} (\frac{-i g_{\nu\beta}}{p^2})
\end{eqnarray}
where,
\begin{eqnarray}
     \Pi^{\mu \nu}&=& ie^2 \int \frac{d^4 k}{(2\pi)^4} \frac{Tr\left[\gamma^{\nu} (\slashed{k}+m)\gamma^{\mu}(\slashed{k}-\slashed{p}+m)\right]}{(k^2-m^2)((k-p)^2-m^2)}	     \nonumber\\
     &=& ie^2\int_{0}^{1}dx 8x(1-x) (p^{\mu}p^{\nu}-g^{\mu\nu}p^2)\int\frac{d^4\ell}{(2\pi)^4}\frac{1}{(\ell^2-\Delta)^2}\nonumber\\&+&ie^2\int_{0}^{1}dx\int\frac{d^4\ell}{(2\pi)^4}4\Big(\frac{2\ell^{\mu}\ell^{\nu}}{(\ell^2-\Delta)^2}-\frac{g^{\mu\nu}}{(\ell^2-\Delta)}\Big).
\end{eqnarray}
In order to check the gauge invariance of this amplitude, it is better to separate the polarisation tensor in the gauge invariant and non-invariant terms as follows
	\begin{eqnarray}
	    \Pi^{\mu \nu}   &=& \underbrace{ie^2\int_{0}^{1}dx 8x(1-x) (p^{\mu}p^{\nu}-g^{\mu\nu}p^2)\int\frac{d^4\ell}{(2\pi)^4}\frac{1}{(\ell^2-\Delta)^2}}_{\text{gauge invariant term}}+\underbrace{4ie^2\int_{0}^{1}dx (2 I_2^{\mu\nu}-I_2g^{\mu\nu})}_{\text{gauge non-invariant term}}
       \label{eqn_pimunu}
	\end{eqnarray}
  where, $I_2^{\mu\nu}$ and $I_2$ are the quadratically divergent irreducible loop integrals.  Note that we can always express Feynman integrals into a set of irreducible loop integrals.

Now, the requirement of gauge invariance of this amplitude demands the second term to be zero in any regularisation scheme. Consequently, gauge invariance forces the two integrals in the last terms to have a relation\footnote{Similar relations were obtained in \cite{Wu:2003dd} using gluon vacuum polarisation amplitude. Using those relations the validity of the symmetry preserving loop regularisation scheme is established.} given by
\begin{equation}
   2I_2^{R \mu\nu}=I_2^Rg^{\mu\nu}
   \label{rel_quad}
\end{equation}
where, $I_2^{R\mu\nu}$ and $I_2^R$ are the regularised loop integrals with quadratic divergences. In Den. Reg., they are given by (from now on, we do not explicitly show the factors of the fictitious scale $\mu$ which can be reintroduced when required) 
\begin{eqnarray}
	   I_2^R &=&f_{(1,2)}\int \frac{d^4l}{(2\pi)^4} \frac{1}{(l^2-\Delta)^{1+\epsilon}}\nonumber\\
	    &=&\frac{-i(-1)^{\epsilon}}{16\pi^2}f_{(1,2)}\frac{\Gamma(\epsilon-1)}{\Gamma(1+\epsilon)}\Big(\frac{1}{\Delta}\Big)^{\epsilon-1},
	    \label{eqn_i3}
	\end{eqnarray}
	and 
		\begin{eqnarray}
	   I_2^{R\mu\nu} &=&f_{(2,2)}\int \frac{d^4l}{(2\pi)^4} \frac{l^{\mu}l^{\nu}}{(l^2-\Delta)^{2+\epsilon}}\nonumber\\
	    &=&\frac{-i(-1)^{\epsilon}}{16\pi^2}f_{(2,2)}\frac{\Gamma(\epsilon-1)}{\Gamma(2+\epsilon)}\Big(\frac{1}{\Delta}\Big)^{\epsilon-1},
	    \label{eqn_i4}
	\end{eqnarray}
	respectively. {From the computation of these integrals and taking the $m\rightarrow 0$ limit, the quantities $f_{(1,2)}$ and $f_{(2,2)}$ turn out to be equal to $(1-\epsilon) \Gamma^{-2}(1-\epsilon)$. Here, the factor $(1-\epsilon)$ cancels the quadratic UV divergences in both the integrals as discussed in \cite{Horowitz:2022uak}. This equality between $f_{(1,2)}$ and $f_{(2,2)}$ is very crucial for Eqn.(\ref{rel_quad}) to hold true in denominator regularisation.} 
	
	Similarly, demanding the gauge invariance for the gluon vacuum polarisation results in a similar relation for the regularised logarithmically divergent integrals which reads as\footnote{This relation is easy to see in the $R_{\xi}$ gauge.}
\begin{equation}
    4I_0^{R \mu\nu}=I_0^R g^{\mu\nu}
    \label{rel_log}
\end{equation}
where, $I_0^{R\mu\nu}=f_{(3,0)}\int \frac{d^4l}{(2\pi)^4} \frac{l^{\mu}l^{\nu}}{(l^2-\Delta)^{3+\epsilon}}$, and $I_0^R =f_{(2,0)}\int \frac{d^4l}{(2\pi)^4} \frac{1}{(l^2-\Delta)^{2+\epsilon}}$ are the regularised logarithmically divergent loop integrals in denominator regularisation. Eqn.(\ref{rel_log}) holds true in Den. Reg. for $f_{(3,0)}=f_{(2,0)}$ which is also consistent  with the results in \cite{Horowitz:2022uak}. 

The relation in Eqn.(\ref{rel_log}) plays a very crucial role in determining the amplitude of $H\to\gamma\gamma$ which can be seen by expressing Eqn.(\ref{eqn_imunuf}) in terms of the irreducible loop integrals such that
	\begin{eqnarray}
	    I^{\mu\nu}&=& \int \frac{d^4l}{(2\pi)^4} \left[\frac{4l^\mu l^\nu}{(l^2-\Delta)^3} - \frac{g^{\mu\nu}}{(l^2-\Delta)^2} - \frac{g^{\mu\nu}\Delta }{(l^2-\Delta)^3}\right]\nonumber\\
	    &=&4I_0^{\mu\nu}-I_0 g^{\mu\nu}-\int \frac{d^4l}{(2\pi)^4} \frac{g^{\mu\nu}\Delta }{(l^2-\Delta)^3}.
	    \label{eqn_i1}
	\end{eqnarray}
Now using Den. Reg. and the relation in Eqn.(\ref{rel_log}), we obtain

	\begin{eqnarray}
	    I^{\mu\nu}&=&-\int \frac{d^4l}{(2\pi)^4} \frac{g^{\mu\nu}\Delta }{(l^2-\Delta)^{3+\epsilon}}\nonumber\\
	    &=&\frac{i}{32\pi^2}g^{\mu\nu}
	    \label{eqn_i2}
	\end{eqnarray}
which is the desired result and matches well with the one obtained from dimensional regularization. Moreover, we have also checked how the scale-less integrals fare in Den. Reg. 
\begin{equation}
    I_{sl} = \int \frac{d^4k}{(2\pi)^4} \frac{1}{(k^2)^\alpha} \underset{\text{Den. Reg.}}{\xrightarrow{\hspace*{1.2cm}}} f_{(\alpha,4-2\alpha)}\int \frac{d^4k}{(2\pi)^4} \frac{1}{(k^2)^{\alpha+\epsilon}}.
\end{equation}
Upon performing these integrals in the Den. Reg. and employing analytic continuation, these integrals are found to vanish. We thus have the result 
\begin{equation}
    I_{sl} = \int \frac{d^4k}{(2\pi)^4} \Bigg\lbrace  \frac{1}{(k^2)^{\alpha}},(k^2)^{\alpha}, \ldots, k^{\alpha_1}k^{\alpha_2}...k^{\alpha_n}\Biggr\rbrace=0.
\end{equation}
This result also turns out to be the same as in dimensional regularisation.

It is clear that operationally i.e., at the level of integrals, Den. Reg. comes very close to dimensional regularisation. There is one crucial difference though. In Den. Reg., $\epsilon$ appears as  the power of the combined denominator and thus unlike dimensional regularisation, also appears in the argument of the gamma function (eg. $\Gamma(1+\epsilon)$ or $\Gamma(2+\epsilon)$ in Eqn.(13) and Eqn.(14), respectively) in the denominator of the final expression of the integrals. The expansion around $\epsilon=0$ then produces extra $\mathcal{O}(\epsilon)$ and higher terms. These in principle may lead to different finite terms. This would then have to be viewed as part of the eventual subtraction scheme if Den. Reg. is adopted. When done systematically and consistently, there would not be any ambiguities in physical results. 

\section{Conclusion and Discussions}
\label{discussion}
In this work, we have tried to look at the generality and applicability of the recently proposed Den. Reg. scheme to other processes with the help of Higgs decay to two photon process. With a very naive application of Den. Reg. to $H\to\gamma\gamma$, this prescription seemed to fail in providing us the physical answer as the two integrals seem to simply cancel against each other. Such a cancellation would have dramatic effects, not only on the numerical value of the $H\to\gamma\gamma$ rate but also on the decoupling of W-boson contribution. For purely scalar integrals, say the scalar part in Eq. (2), the divergence is regularised as:
\begin{align*}
    & \text{Dimensional Regularisation: } (4-\epsilon) +2-3 \\
    & \text{Denominator Regularisation: } 4+2-(3+\epsilon)
\end{align*}
Thus, they would yield the same results, up to possible extra constant factors arising upon expanding around $\epsilon=0$ which then would be part of that scheme. But when faced with divergent integrals such as Eqn.(\ref{eqn_int}), denominator regularization does not seem to distinguish between the two terms present in Eqn.(\ref{eqn_int}) and leads to zero. This, fortunately, turns out to be a hasty and erroneous conclusion. 

For a detailed analysis on this issue, we then first revisited the QED vacuum polarisation in this scheme and tried to look for some interesting features, if any. The requirement of the gauge invariant answer for QED vacuum polarisation results in not very surprising but an interesting relation between the quadratically divergent integrals as given in Eqn.(\ref{rel_quad}). Similar relations are found for the logarithmically divergent integrals by demanding gauge invariance for the gluon vacuum polarisation (see Eqn.(\ref{rel_log}))\footnote{Such, or similar, relations would be expected in any acceptable regularization scheme once gauge invariant results are demanded. More such relations may be obtained by checking Ward identities related to other symmetries. At higher loops or for more involved Feynman amplitudes, one might uncover a larger set of such relations.}. Upon realising these interesting relations between the two integrals and applying them to $H\to\gamma\gamma$, we obtained the desired results. Consequently, we conclude that Den. Reg. turns out to be a consistent regularisation scheme by consistently using the relations which appeared as a consequence of gauge invariance. The scale-less integrals are also found to provide the same results  as dimensional regularisation.

In the future, it will be highly interesting to perform more tests involving higher orders, and loops and check the validity of this scheme along with the hunt for more interesting features hidden, if any. A particularly important aspect will be to check how universal the quantities $f_{(n,p)}$ are. It seems that for a given amplitude, two $f_{(n,p)}$'s with the same $p$ but different $n$ are the same, at least when gauge invariance is demanded. It remains to be checked if this continues to be the case for more general cases. Moreover, looking at practical applications, processes involving $\gamma_5$ will be highly interesting since such problems face potential ambiguities in dimensional regularisation. Den. Reg. being a 4-dimensional scheme is expected to be free from any such ambiguities and thus could provide more consistent answers for such applications. Another immediate and important application would be to infrared divergences. Since Den. Reg. seems to share a lot of practical similarities with dimensional regularisation, once a clear prescription for it is established, it may provide a powerful alternative to dimensional regularisation and be free of some issues encountered with dimensional regularisation.

\bibliography{denreg}{}
\bibliographystyle{unsrt}
\end{document}